\documentclass{ceurart}

\usepackage{listings}
\begin{document}
\copyrightyear{2025}
\copyrightclause{Copyright for this paper by its authors.
  Use permitted under Creative Commons License Attribution 4.0
  International (CC BY 4.0).}

\conference{TDI 2025: 3rd International Workshop on Trends in Digital Identity, February 3, 2025, Bologna, Italy}



\title{did:self A registry-less DID method}

\author{Nikos Fotiou}[%
email=fotiou@excid.io
]
\cormark[1]

\author{George C. Polyzos}[%
email=polyzos@excid.io,
]

\author{Vasilios A. Siris}[%
email=siris@excid.io,
]

\address{ExcID, 113 62, Athens, Greece}
\cortext[1]{Corresponding author.}
\fntext[1]{All authors contributed equally.}

\begin{abstract}
We introduce \texttt{did:self}, a Decentralized Identifier (DID) method that does not depend on any trusted registry for storing the corresponding DID documents. Information for authenticating a \texttt{did:self} \emph{subject} can be disseminated using any means and without making any security assumption about the delivery method. \texttt{did:self} is lightweight, it allows controlled delegation, it offers increased security and privacy, and it can be used for identifying people, content, as well as IoT devices. Furthermore, DID documents in \texttt{did:self} can be implicit, allowing re-construction of DID documents based on other authentication material, such as JSON Web Tokens and X.509 certificates. 
\end{abstract}


\maketitle

\section{Introduction}
In the age of ``surveillance capitalism''\footnote{https://en.wikipedia.org/wiki/Surveillance\_capitalism} personal data is massively collected by large corporations that prioritize profit over privacy.  At the same time, the alarming number of large scale data breaches, such as the Equifax data  breach,\footnote{https://en.wikipedia.org/wiki/2017\_Equifax\_data\_breach} and incidents of massive disclosure of personal data, such as the Facebook-Cambridge Analytica  scandal,\footnote{https://en.wikipedia.org/wiki/Facebook–Cambridge\_Analytica\_data\_scandal} is a clear indication of the lack of proper security and privacy safeguards.  To address these concerns, a significant number of efforts advocating decentralization has sprung up. W3C's \emph{Decentralized Identifiers} (DIDs) is a notable example of such an effort in the domain of identification. W3C's DID specifications pursue a portable digital  identity that does not depend on a centralized authority and  can never be taken away by third-parties~\cite{did-primer}. Decentralized identification systems are also beneficial to service providers, many of which collect personally identifying and private information to support their personalized services. A decentralized, user-centric approach for handling identity will alleviate service providers from the burden of collecting sensitive data~\cite{Tot2019} for which they are liable.      

A DID is a new type of self-administered, globally unique identifier, which is \emph{resolvable} and \emph{cryptographically verifiable}~\cite{did-primer}. In particular, a DID is resolved to a \emph{DID document} that contains (among other things) cryptographic material that can be used for verifying DID ownership (e.g., a public key). DID documents are usually stored and maintained by a \emph{registry}~\cite{did-reg}.  Registries may be administrated by a single entity or a consortium of trusted entities (e.g., a Web server or a permissioned blockchain), or they may be provided by public systems such as a public, permissionless blockchain (e.g., Ethereum). Registries of the former type require some level of trust and  may introduce security and privacy risks. On the other hand, the interaction with  blockchain registries may involve significant computational overhead (which may not be  tolerable by an IoT device), or even monetary cost. Moreover, in general, existing registries lack interoperability and in many cases the information associated with a DID cannot be transferred from one registry to another. With these in mind, we propose \texttt{did:self}, a new DID \emph{method}, which is compatible with W3C specifications and at the same time  does not depend on any type of registry.   

Identifiers in \texttt{did:self} are created using the thumbprint of a public key. Accordingly, DID documents are protected by a \emph{proof} generated by the DID controller; this proof can be validated with the public key that was used for creating the corresponding identifier. Therefore, given a \texttt{did:self} DID and  the corresponding DID document and proof, any entity can verify their binding, as well as the correctness and the integrity of the DID document. Similarly to self-signed digital certificates, proof verification does not require any auxiliary information. A unique feature of \texttt{did:self} is that it supports secure DID sharing and controlled delegation. By leveraging this property, earlier versions of our DID method were used for implementing content-centric security for Information-Centric Networking architectures~\cite{Fot2021b,Fot2023}, for securing IoT group communication~\cite{Fot2022}, as well as for improving the security of the Inter-Planetary File System (IPFS)~\cite{Fot2021c}.

In this paper, we introduce an update to the \texttt{did:self} specification that brings the following improvements:
\begin{itemize} 
    \item DIDs can have a (human-readable) suffix that provides clearer semantics about the entities that share the same identifier. 
    \item We generalize the concept of proof and  introduce implicit DID documents. To this end, we propose two methods for generating DID documents using JSON Web Tokens and X.509 certificates. This enables interoperability with existing  authentication and authorization systems.
\end{itemize}

The remainder of this paper is organized as follows. In Section~2 we introduce DIDs. In Section~3 we present the design of our solution. In Section~4 we discuss related work in this area. We conclude our paper in Section~5.

\section{Background: W3C Decentralized Identifiers}

\label{sec:back}
Decentralized Identifiers (DIDs), defined by W3C, are a new type of globally unique identifier 
designed to enable individuals and organizations to generate their own identifiers using systems they trust.
A DID architecture can be regarded as a key-value lookup system, where the key is 
the Decentralized Identifier (DID) and the value is a DID \emph{document}.
DID specifications allow multiple DID \emph{methods} to co-exist; the main differentiating
factor among existing DID methods is how the resolution from a DID to the corresponding
DID document is implemented. A DID is expressed as a URI composed of three parts
separated by ``:''; the first part is the word ``did'', the second part is the
DID method name, and the third part is a method specific identifier. 

A DID document contains ``properties'', serialized according to a particular syntax,
that may include (among other things) \emph{verification methods} (e.g., 
public keys) and \emph{verification relationships} that express a relationship
between a verification method and the DID \emph{subject}. Examples of verification
relationships are \emph{authentication}, which means that
a verification method is used for authenticating the DID subject, and
\emph{assertion}, which means that a verification method is used for verifying assertions, e.g., digital signatures, made by the DID subject. 

Listing~\ref{list:did} illustrates an example of a DID document. The
DID of the subject is \texttt{did:self:iQ9PsBKOH1nLT9FyhsUGvXyKoW00yqm\_-\_rVa3W7Cl0/device1} (line 2). 
In lines 3-10 a verification method is defined, which is a 
public key. This public key has an 
identifier (i.e., ``\#key1''), and it is encoded using the type 
``JsonWebKey2020''~\cite{did-reg}, which is an encoding based on 
RFC 7517 JSON Web Keys.
Next, two verification relationship are defined, i.e., ``authentication'' and ``assertion''
which include the identifier of the defined key. Therefore, in this example ``\#key1''
can be  used for authenticating  the DID subject and for verifying digital signatures it generates. 
\begin{minipage}{\linewidth}
\begin{lstlisting} [caption={A sample DID document},label={list:did}]
  {
    ``id": ``did:self:iQ9PsBKOH1nLT9FyhsUGvXyKoW00yqm_-_rVa3W7Cl0/device1",
    ``verificationMethod": [{
      ``id": ``#key1",
      ``type": ``JsonWebKey2020", 
      ``publicKeyJwk": {
        ``kty": ``EC",
        ``crv": ``P-256",
        ``x": ``YOGmYaMKzwTFytWHN2hGC-2VpPqGqj_sDSckB2IvCgI",
        ``y": ``7iWuiXQlLXvROjdMA2WNHhGz0jxu6u41n83YupNteo"
      }
      }
    ],
    ``authentication": [``#key1"],
    ``assertion": [``#key1"],
  }
\end{lstlisting} 
\end{minipage} 
\section{The did:self DID method}
A \texttt{did:self} DID\footnote{In the following we use the terms \texttt{did:self} DID and \texttt{did:self} identifier interchangeably} is generated by a \emph{controller} who creates a public/private key pair. The \texttt{did:self} identifier is the thumbprint of the JSON Web Key (JWK) representation of the controller's public key (as defined by RFC 7638~\cite{rfc7638}), optionally followed by a suffix. Therefore a \texttt{did:self} identifier has the following form:

\[did{:}self{:<}thumbprint{>/<}suffix{>}\]

\begin{figure}
    \includegraphics[width=0.9\linewidth]{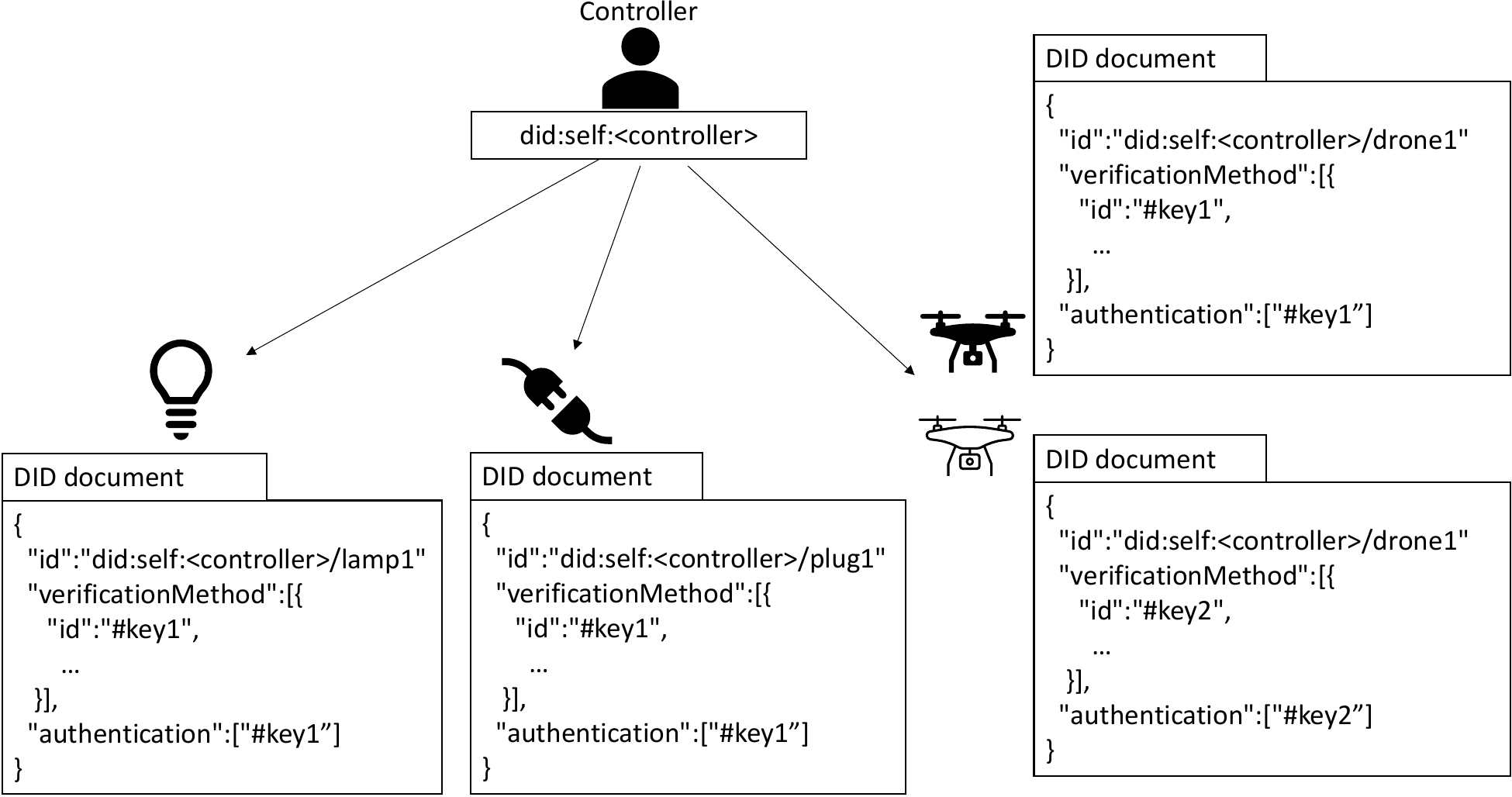}
    \caption {An IoT use case where did:self is applied. It can be observed that all IoT devices share a DID with the same prefix. Furthermore, the two drones have the same DID.}
    \label{fig:usecase}
\end{figure}

The corresponding DID document is protected by a \emph{proof} generated using the controller's private key. A  controller is not necessarily the DID \emph{holder}. Our method supports multiple holders per DID, each of which may use a public/private key pair it controls for implementing a \emph{verification relationship} (e.g., \emph{authentication}). As a motivating example consider the case of an IoT system, where the IoT system owner is the controller and each IoT device can be a holder. IoT devices can use a key pair they control to implement a verification relationship. In this case, for each IoT device a DID document is created that defines a verification relationship based on the public key that the device controls (see for example Listing~\ref{list:did}). The proof for this document is generated by the controller. In this use case, for each IoT device a suffix can be added to the DID. Additionally, two IoT devices may share the same DID, allowing, for example, device rotation in case of failure or representation of virtual devices composed of multiple physical ones (e.g., ``smart home''). This use case is illustrated in Figure~\ref{fig:usecase}, which shows two drones that share the same DID, but use a different authentication key. 

A straightforward approach for creating a proof is by generating a JSON Web Signature (JWS)~\cite{rfc7515} using the private key of the controller's key pair. In that case, the JWS header must include the \texttt{jwk} field, which is used as follows:

\begin{itemize}
 \item \texttt{jwk} The JWK representation of the controller's public key, which can be used for verifying the proof. The thumbprint of this key must match the  thumbprint included in the \texttt{did:self} identifier.
\end{itemize}
The payload of the JWS is the DID document. Therefore, validating such a proof is a simple three-steps process: (i) extract the public key from the JWS header, (ii) verify that the thumbprint of the extracted public key is equal to the thumbprint included in the  \texttt{did:self} identifier, and (iii) validate the JWS signature using as input the extracted public key and the DID document. 

In the next sections  we discuss how \texttt{did:self} can be used with JSON Web Tokens (JWT) and X.509 certificates. 

\subsection{Representing DID documents as JSON Web Tokens}
A \texttt{did:self} DID document that includes a single verification method can be represented as a JSON Web Token (JWT)~\cite{rfc7519}. The following table specifies how the payload claims of such a JWT are created:

\begin{center}
\begin{table}[h!]
\begin{tabular}{ |l|l| } 
 \hline
 \textbf{Claim} & \textbf{Value}  \\ 
 \hline
 \texttt{iss} & The \texttt{did:self} identifier without any suffix\\
 \texttt{sub} & The suffix of the \texttt{did:self} identifier\\
 \texttt{cnf} & A public key controlled by the DID holder represented as a JWK\\
 \hline
\end{tabular}
\caption{The claims of a JWT representing the DID document of a did:self DID.}
\label{tab:relationships}
\end{table}
\end{center}

Listing~\ref{list:jwt} includes the payload of a JWT representing the DID document of Listing~\ref{list:did}. Such a JWT is signed by the controller using its private key. The JWS header must include either the \texttt{jwk} claim whose value will be the JWK representation of the controller's public key or the \texttt{x5c} field representing a certificate chain (see next section). Again, the JWT verification is a three-steps process: (i) extract the public key from the JWS header, (ii) verify that the thumbprint of the extracted public key is equal to the thumbprint included in the  \texttt{iss} claim, and (iii) validate the JWS signature using as input the extracted public key and the JWT.

\begin{minipage}{\linewidth}
\begin{lstlisting} [caption={A did:self DID document represented as a JWT},label={list:jwt}]
  {
    ``iss": ``did:self:iQ9PsBKOH1nLT9FyhsUGvXyKoW00yqm_-_rVa3W7Cl0",
    ``sub": ``device1",
    ``cnf": {
      ``jwk": {
        ``kty": ``EC",
        ``crv": ``P-256",
        ``x": ``YOGmYaMKzwTFytWHN2hGC-2VpPqGqj_sDSckB2IvCgI",
        ``y": ``7iWuiXQlLXvROjdMA2WNHhGz0jxu6u41n83YupNteo"
       }
      }
  }
\end{lstlisting} 
\end{minipage}

Using this approach \texttt{did:self} can be straightforwardly used with protocols such as  Demonstrating Proof-of-Possession at the Application Layer (DPoP) of OAuth 2.0~\cite{dpop}. 

\subsection{Representing DID documents as X.509 certificates}
Similarly, a \texttt{did:self} DID document that includes a single verification method can be represented as an X.509 certificate chain. The first certificate of the chain is a self-signed certificate that includes the controller's public key and the \texttt{did:self} identifier (without any suffix) in the \texttt{Subject Alternative Name} field. This certificate is signed using the controller's private key. The last certificate of the chain includes the holder's public key and the \texttt{did:self} identifier (with a suffix if necessary) in the \texttt{Subject Alternative Name} field. The chain may include intermediate certificates issued by the DID controller. The intermediate certificates give greater flexibility to the controller, which can have multiple public keys associated with its DID with varying level of security. The validation of the certificate chain follows the standard X.509 validation rules. Additionally, a verifier should verify that the thumbprint included in the \texttt{Subject Alternative Name} is the same as the thumbprint of the JWK representation of the public key included in the first certificate of the chain. 

\section{Related work}
\label{sec:back}
Although DID document registries management, they create lock-in conditions, decrease DID self-sovereignty, and add communication overhead, which in some scenarios is intolerable. For this reason, various efforts propose registry-less DID systems.
NaCL:did~\cite{nac2020} and did:key~\cite{Lon2020} are two such systems that use (the hash of) a public
key as the DID.
Similarly, ethr-did~\cite{eth2021} uses Ethereum addresses as DIDs. 
The main drawback of these systems is that they support a single verification method,
which is the public key from which the DID has been derived. For this reason, they
cannot not support real DID documents; instead, in these methods DID documents are implied. In contrast, did:self
supports DID documents, which may include multiple verification methods.  

did:peer~\cite{Dev2020} is a registry-less DID system, which
allows DID documents. 
DID document modifications are encoded in authenticatable \emph{deltas}. A DID document can be
modified by multiple entities, and for this reason did:peer relies on a \emph{conflict-free replicated 
data type} for consolidating all deltas and eventually creating the final DID document. This approach
has some side-effects: it makes DID document reconstruction complex and it poses restrictions on the
format of the document. Our approach is much simpler, since the only additional operation required for
retrieving a DID document is the validation of at most two digital signatures. Furthermore,  our
approach does not pose any restriction to the fields a DID document 
may contain, therefore, our solution is 
sustainable and future proof; this is very important considering that DID
specifications have not yet been finalized. 

did:web~\cite{Zag2020} is a registry-based DID system that allows users to use
any Web server as a registry. With did:web, DIDs are bound to the URL of the Web server and
DID document resolution is trivially implemented using HTTPS. did:web offers a higher degree of 
sovereignty compared to 
many existing systems, but still relies on the security of HTTPS for securely 
distributing DID documents. On contrary, did:self does not need any
secure communication channel for distributing DID documents. 

ION~\cite{dif2020} is also a registry-based DID system that has been adopted
by Microsoft.\footnote{https://techcommunity.microsoft.com/t5/identity-standards-blog/ion-we-have-liftoff/ba-p/1441555}
In ION, DID documents are stored in a ``Content-addressable storage network'' 
(e.g. the InterPlanetary File System (IPFS)~\cite{Ben2014}). Modifications
to the DID documents are stored in ``delta'' files, which are eventually
``anchored'' in a decentralized sequencing oracle (ION uses Bitcoin for this
purpose). Therefore, DID documents are reconstructed by retrieving the appropriate
``delta'' files, following the ``pointers'' stored in the sequence oracle. 
Compared to did:web, ION relies on a more distributed trust model, but at the
same time adds complexity and in some cases monetary cost.


The Key Event Receipt Infrastructure (KERI) proposal \cite{Smi2019} considers as 
the primary root-of-trust self-certifying identifiers that are strongly bound at 
issuance to a cryptographic signing (public, private) key-pair, similar to our proposal. 
Key rotation is 
performed with a signed transfer statement. The KERI proposal considers an indirect 
mode 
where a set of witnesses (replicas) 
store and forward key events 
to any requester/validator. Although the existence of witnesses enhances fault tolerance 
and availability, it entails additional requirements and complexity related to witness 
designation and policy, witness consensus, and handling of out-of-order events. did:self 
avoids such complexity by focusing on use cases where it is sufficient to disseminate 
the corresponding DID document through any method, even between untrusted parties.

Compared to all these DID methods, did:self is the only DID method that allows
multiple DID documents per DID, as well as controlled DID document delegation. As we discuss
in Section 4, these properties enable intriguing security and privacy features. 

Due to the 
security and privacy characteristics of the DID paradigm,
many research efforts investigate the potential of using DIDs to improve the security and privacy of emerging
technologies. Chadwick et al.~\cite{Cha2019} propose the integration of DIDs and
Verifiable Credentials with the FIDO Universal Authentication Framework (UAF) in order
to provide safer and more private online account management. 
DIDs are also considered for improving the security and privacy of IoT
systems (e.g.,~\cite{Ans2019},\cite{Fed2020},\cite{Kor2019},\cite{Ter2020}). Davie et al.~\cite{Dav2019} propose
a four-layer architectural stack, based on DIDs, to establish trust between peers over 
the Internet and other digital networks. Lagutin et al.~\cite{Lag2019} investigate
the application of DIDs and Verifiable Credentials in the OAuth 2.0 authorization process. 
Finally, Munoz~\cite{Mun2019} discusses the
advantages of integrating eIDAS and DIDs. Our work is complementary to these
approaches: being standard compliant, did:self could be the DID technology that
these proposed systems may eventually use.

\section{Conclusion}
In this paper we presented the \texttt{did:self} DID method and introduced two new extensions: identifier suffixes and implicit DID documents. 
Our DID method gives controllers absolute control over their identifiers and removes any dependency on middlemen, hence increasing self-sovereignty and privacy. At the same time, \texttt{did:self} allows multiple DID documents per DID enabling delegation and sharing. Our solution is not only compatible with W3C specifications, but  is also easier to integrate in existing systems, by removing the complexity of interacting with a DID registry. 

By leveraging implicit DID documents, we presented two constructions for generating a \texttt{did:self} DID document: one using JSON Web Tokens and another based on X.509 certificates. This enables interoperability with existing authentication and authorization systems. At the same time however, it prevents taking advantage of the full potential of DIDs, since legacy DID documents can carry more information beyond a single public key. 

Although the lack of a registry and the support for multiple DID documents per
DID enable novel applications and offer significant security and privacy advantages,
they create challenges related to revocation. Currently, \texttt{did:self} replies mainly
on the expiration time to handle revoked verification material, but other
more efficient mechanisms are also being explored.

\bibliography{biobliography.bib}

\end{document}